\documentclass[useAMS,usenatbib]{mnras} 
\usepackage{amsmath}
\usepackage{graphicx}
\usepackage[T1]{fontenc}
\usepackage{epstopdf}

\title[]{The global warming of group satellite galaxies}
\author[C. Yozin and K. Bekki]{C. Yozin\thanks{{\bf E-mail} 21101348@student.uwa.edu.au; kenji.bekki@uwa.edu.au} and K. Bekki\footnotemark[1]
\\\\
ICRAR, M468, The University of Western Australia, 
35 Stirling Highway, Crawley
Western Australia, 6009, Australia}

\begin{document}
\date{Accepted 2015 January. Received 2015 January; in original form 2015 January}
\pagerange{\pageref{firstpage}--\pageref{lastpage}} \pubyear{2013}
\maketitle
\label{firstpage}


\newcommand{\asym}[1]{$<$A$_{\rm #1}$$>$}
\newcommand{\atlas}{{\sc ATLAS}$^{\rm 3D}$}
\newcommand{\bia}{$b/a$}
\newcommand{\bt}{B/T}
\newcommand{\cpet}{c$_{\rm 90/50}$}
\newcommand{\cnfw}{c$_{\rm NFW}$}
\newcommand{\ex}[1]{10$^{\rm #1}$}
\newcommand{\hi}{H~{\sc I}}
\newcommand{\ih}{$h^{\rm -1}$}
\newcommand{\fgas}{f$_{\rm gas}$}
\newcommand{\kms}{kms$^{\rm -1}$}
\newcommand{\lcdm}{$\Lambda$CDM}
\newcommand{\lr}{$\lambda_{\rm Re}$}
\newcommand{\lre}{$\lambda_{\rm Re}$-$\epsilon_{\rm e}$}
\newcommand{\lsol}{L$^*$}
\newcommand{\maga}{mag arcsec$^{\rm -2}$}
\newcommand{\mstar}{M$_{\ast}$}
\newcommand{\mex}[2]{#1$\times$10$^{\rm #2}$}
\newcommand{\mhms}{M$_{\rm h}$/M$_{\ast}$}
\newcommand{\msol}{M$_{\odot}$}
\newcommand{\muc}{$\mu$(g,0)}
\newcommand{\mur}{$\mu$(g,r)}
\newcommand{\mvir}{M$_{\rm vir}$}
\newcommand{\pcsq}{pc$^{\rm -2}$}
\newcommand{\reff}{r$_{\rm e}$}
\newcommand{\rs}{r$_{\rm s}$}
\newcommand{\rvir}{r$_{\rm vir}$}
\newcommand{\rvirmw}{r$_{\rm vir,MW}$}
\newcommand{\vrot}{V$_{\rm rot}$}
\newcommand{\z}[1]{$z=#1$}
\DeclareRobustCommand{\VAN}[3]{#2}

\begin{abstract}

Recent studies adopting $\lambda_{\rm Re}$, a proxy for specific angular momentum, have highlighted how early-type galaxies (ETGs) are composed of two kinematical classes for which distinct formation mechanisms can be inferred. With upcoming surveys expected to obtain $\lambda_{\rm Re}$ from a broad range of environments (e.g. SAMI, MaNGA), we investigate in this numerical study how the $\lambda_{\rm Re}$-$\epsilon_{\rm e}$ distribution of fast-rotating dwarf satellite galaxies reflects their evolutionary state. By combining N-body/SPH simulations of progenitor disc galaxies (stellar mass $\simeq$10$^{\rm 9}$ M$_{\odot}$), their cosmologically-motivated sub-halo infall history and a characteristic group orbit/potential, we demonstrate the evolution of a satellite ETG population driven by tidal interactions (e.g. harassment). As a general result, these satellites remain intrinsically fast-rotating oblate stellar systems since their infall as early as $z=2$; mis-identifications as slow rotators often arise due to a bar/spiral lifecycle which plays an integral role in their evolution. Despite the idealistic nature of its construction, our mock $\lambda_{\rm Re}$-$\epsilon_{\rm e}$ distribution at $z<0.1$ reproduces its observational counterpart from the {\sc ATLAS}$^{\rm 3D}$/SAURON projects. We predict therefore how the observed $\lambda_{\rm Re}$-$\epsilon_{\rm e}$ distribution of a group evolves according to these ensemble tidal interactions.

\end{abstract}


\section{Introduction}


In a major step beyond classifying early-type galaxies (ETGs) according to their morphology alone, the spatially-resolved velocity fields arising from recent volume-limited IFS (integral-field spectroscopy) surveys (SAURON, \atlas{}, CALIFA) have provided deeper insight into their formation. By consolidating this kinematic data into a proxy ($\lambda_{\rm Re}$) for the specific angular momentum, a metric tightly correlated with morphology \citep{obre14}, it has become possible to distinguish two classes of ETGs: fast and slow rotators \citep[FR and SR respectively;][]{emse07}. The pressure-supported SRs are predominantly round, old and massive (stellar mass \mstar{}$>$\ex{10.5}{} \msol{}), whilst FRs appear continuous (in both kinematics and shape) with late-type galaxies \citep[LTGs;][]{emse11, kraj11, weij14}.

This dichotomy has parallels with the red-blue classification of galaxies, and likewise, SR ETGs are conceived to be the natural products of their hierarchical assembly \citep{khoc06}. Myriad numerical studies have addressed this scenario; those recently motivated by \atlas{} have established that both binary major mergers and successive minor mergers can reproduce the kinematic peculiarities of SRs \citep{bour07, jess09, bois11, mood14}. In practice, the cosmological assembly of a SR galaxy follows a non-trivial evolution in $\lambda_{\rm R}$ depending on merger mass ratios/gas fractions \citep{naab14}.

As IFS surveys broaden their sample size, more focus will shift to the influence of different environments on forming SR/FRs. Early results in this context suggest SRs are preferentially located in dense locations \citep{capp13}, with a clear radial trend in ellipticity among clusters \citep{deug15}. Mounting evidence also suggests the formation of FR ETGs \citep[such as S0s, the underlying driver of the morphology-density relation;][]{dres80} most commonly follows the transformation of LTGs in groups \citep{wilm09, just10}. 

The distinct kinematics between LTGs and S0s is a strong argument against fading alone in this transition, although it remains to be established which of the associated group mechanisms dominate. In a precursor to this paper, \citet{bekk11} demonstrated that group tides/harassment can faciliate spiral-to-S0 transformation, and \citet{foga15} showed how the corresponding evolution in $\lambda_{\rm Re}$ is consistent with those of the pilot SAMI survey. On the other hand, \citet{quer15} proposed major mergers as a viable formation pathway, in the context of a group host, for those S0s observed in the CALIFA survey. 


To address this issue (and build upon the success of adopting $\lambda_{\rm Re}$ as a discriminant of formation mechanisms), we introduce in this paper a new concept. Based on the premise that environmental mechanisms acting within a group or cluster can be imprinted on the \lre{} distribution of its satellite galaxies, we describe here the appearance of this {\it global warming} from LTGs to ETGs as a function of redshift.

Although tested against the SAURON and \atlas{} samples (the latter comprising 260 galaxies), the work is orchestrated for the purpose of comparison with the upcoming Sydney-AA0 Multi-object IFS survey \citep[SAMI;][]{croo12}. The complete survey will catalogue 3400 galaxies in the local universe ($z$$<$0.1), with excellent coverage anticipated for galaxies of stellar mass \mstar{}$=$\ex{9-10}{} \msol{} in group hosts of dynamical mass \ex{13}{} \msol{}. This particular combination incorporates some of the most common galaxies in the universe \citep{wein09}, the most dominant environment to shape cosmic SFH since \z{1}{} \citep{pope15}, and is therefore the focus in this present study. Additionally, this builds upon our previous work \citep[][hereafter YB15]{yozi15b} concerning the long ($\sim$7 Gyr) quenching/transformation timescales of Magellanic Cloud-type dwarf group-satellites as inferred from near-field observations. 

The paper is organised as follows: in Section 2, we describe the numerical simulations and the post-processing that permits comparison with the \lre{} distributions of \atlas{}/SAMI galaxies; in Section 3, we describe the results of a parameter study with these models, and Section 4 concludes this study with a discussion on the merits of our proposed concept for understanding the dynamical state of a group galaxy population.

\section{Method}

\begin{figure}
\centering
\includegraphics[width=1.\columnwidth]{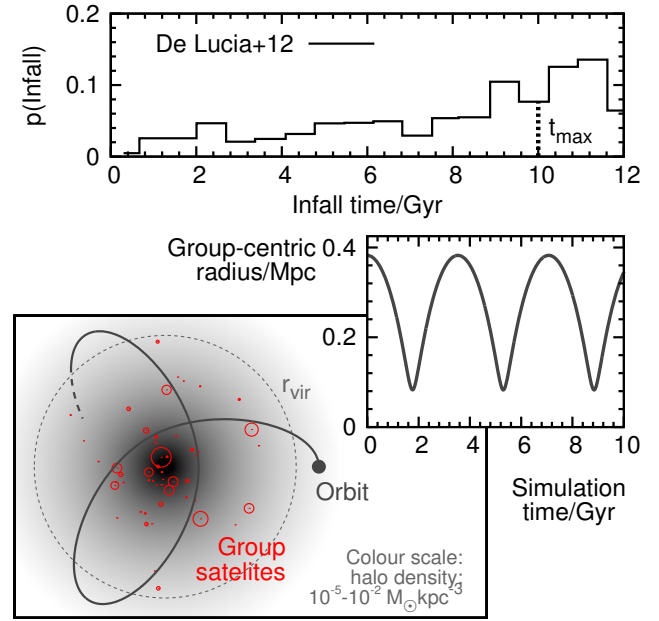}
\caption{(Top panel) Probabilility distribution of lookback times at which a \mstar{}$=$\ex{9}{} \msol{} first becomes a satellite \citep[reproduced from][]{delu12}; (Bottom panel) A schematic of the group model with the virial radius (r$_{\rm vir}$, shown as grey dashed line) bounding the group halo potential (with grey colour scale representing mass density), characteristic galaxy orbit (black solid line) and other satellites (red circles, with size$\propto$mass); (inset panel) The characterstic galaxy orbit, expressed with the group-centric radius as a function of time.}
\end{figure}



Our methodology is comprised of two key steps, the first of which is the simulation of a self-consistent N-Body galaxy model with an interstellar medium (ISM) treated with smoothed particle hydrodynamics (SPH). This model is initially placed on a characteristic orbit within a fixed potential representing the group dark matter halo, as defined in YB15 for satellites with stellar mass (\mstar{}) of \ex{9-9.5}{} \msol{}. 

The second step involves the statistical sampling of this simulated galaxy model evolution, such that we can construct a mock satellite population at a given redshift comprised of galaxies with nominally different infall times. We assume here that infall time ($t_{\rm inf}$) refers to the lookback time at which a satellite first crossed the virial radius of its host halo. Furthermore, our choice of $t_{\rm inf}$ is cosmologically motivated insofar as it is drawn from a probability distribution $p$($t_{\rm inf}$) of satellite infall times (with \mstar{}$\simeq$\ex{9}{} \msol{}; top panel of Fig. 1) corresponding to merger trees constructed from the Millennium Simulation \citep{delu12}. 

The main results of this paper, the mock \lre{} distribution at a given redshift (lookback time $t$), are thus constructed using the Monte-Carlo sampling of $p$($t_{\rm inf}$). Repeated for a minimum of 1000 samplings, we incorporate the simulated galaxy model properties into the mock distribution at a simulation timestep $t_{\rm i} = t_{\rm inf}-t$ if that timestep $t_{\rm i}\geq$0.

These steps are repeated for several galaxy models, differing in properties such as their gas fraction or orbital orientation. This scenario is also compared with those in which other group satellites, represented as point masses, are added to the group model to establish the influence of satellite-satellite tidal interactions. We emphasize therefore that our methodology is based on an idealised group model comprising a fixed dark halo and its satellite population (bottom panel, Fig. 1); the adopted orbit is sufficiently energetic (e.g. the pericentre radius or distance to the group halo centre is sufficiently large) that the explicit modelling of a central galaxy (which by definition lies at the halo centre) is not considered to be necessary.

\subsection{Numerical models}

\begin{table} 
\centering
\caption{A summary of fiducial galaxy model properties}
\begin{tabular}{@{}lc@{}}
\hline
Parameter & Fiducial value \\
\hline
N$^o$. DM particles & \ex{6}{} \\
N$^o$. Stellar particles & 3$\times$\ex{5}{} \\
N$^o$. Gas particles & 2.5$\times$\ex{5}{} \\
Stellar mass/\msol{} (M$_{\rm d}$) & 10$^{\rm 9.25}$ \\
Gas/stellar mass (M$_{\rm g}$/M$_{\rm d}$) & 2 \\
DM halo/stellar mass (M$_{\rm h}$/M$_{\rm d}$) & 85 \\
Stellar disc scalelength (r$_{\rm d}$/kpc) & 1.6 \\
Stellar disc scaleheight (z$_{\rm d}$/kpc) & 0.32 \\
Gas disc scalelength (r$_{\rm g}$/kpc) & 4.2 \\
Gas disc scaleheight (z$_{\rm g}$/kpc) & 0.84 \\
DM Halo virial radius (r$_{\rm vir}$/kpc) & 142 \\
DM Halo concentration ($c$) & 6.5 \\
Metallicity ([Fe/H]) & -0.72 \\
Metallicity gradient ($\Delta$[Fe/H]/kpc) & -0.04 \\
SN expansion timescale ($\tau_{\rm SN}$/yr) & \ex{5.5}{} \\
\hline
\end{tabular}
\end{table} 

We present results from 22 galaxy model simulations, each run with an original parallelised chemodynamical code GRAPE-SPH \citep{bekk09}, combining gravitational dynamics and an ISM with radiative cooling/stellar processing. Our numerical satellite/group models follow that of YB15 closely, to which we refer the interested reader. 

To summarise, the galaxy commences at apocentre on a characteristic orbit (constrained to the $x-y$ plane) which is derived from simulations of collisionless satellite infall with a live model of the dark matter (DM) halo \citep{vill12}. The orientation of the galaxy model is specified by $\theta$ and $\phi$, corresponding to the angle between the $z-$axis and the vector of the angular momentum of the disc, and the azimuthal angle measured from the $x-$axis to the projection of the angular momentum vector of the disc on to the $x-y$ plane, respectively. Our fiducial model adopts $\theta$=45$^{\circ}$ and $\phi$=30$^{\circ}$.

Fig. 1 conveys this orbit and a schematic of the group model which consists of a spherically-symmetric potential alone in the fiducial case (Sections 3.1-2), and incorporates $\sim$50 group satellites (represented in the simulation as point masses) when considering the additional effect of tidal harassment (Section 3.3). 

The group potential has total mass of \ex{13}{} \msol{} and structural properties characteristic of \z{1}{} \citep[virial radius 380 kpc and a NFW density profile with concentration 4.9;][]{vill12}. We use Monte-Carlo sampling to fit the point mass satellites to a Schecter function (slope -1.07) function, limited to a luminosity range 0.01 to 2.5 solar and with the assumption of a mean mass-to-light ratio $\sim$40. Their initial orbits are assigned according to a NFW profile with concentration 3.0.

\subsubsection{Fiducial galaxy model}

Table 1 summarises the principle properties of our fiducial galaxy model, which consists of a gas-rich bulgeless (bulge-to-total ratio, B/T=0) galaxy as in YB15. The initial gas and halo mass with respect to the stellar mass is selected according to best-fit relations from the ALFALFA survey \citep{huan12} and \citet{mill14} respectively, while DM halo properties are obtained from mass-redshift-concentration relations of \citet{muno11}. The stellar disc scalelength r$_{\rm d}$ is observationally motivated from the size-mass relations of \citet{ichi12}, while the gas disc scalelength r$_{\rm g}$ is assumed to be a factor $\sim$3 larger \citep{krav13}; stellar and gas scaleheights z$_{\rm d}$ and z$_{\rm g}$ are assigned as 0.2r$_{\rm d}$ and 2z$_{\rm d}$ respectively. 

While we do not consider the star formation (SF) history here, gas inflow and bulge growth are crucial components to our model evolution, and therefore subgrid SF parametrisations and the efficiency of supernovae (SNe, controlled primarily with the adiabatic expansion timescale $\tau_{\rm SN}$) are constrained according to the Kennicutt-Schmidt law. To offset the idealistic nature of our scenario, we emphasise that our minimum baryonic resolution of $\sim$\ex{4}{} \msol{} is two orders of magnitude less than recent cosmological simulations \citep[e.g. Illustris;][]{gene15}.

\subsubsection{Alternate galaxy models}

The simulations constituting our parameter study are summarised in Table 2. 
\begin{itemize}
\item To account for pre-processing prior to group infall, we construct \lre{} distributions for a galaxy with an initial B/T of 0.2, where the bulge has a Hernquist density profile, isotropic velocity dispersion (with radial velocity dispersion assigned according the Jeans equation of a spherical system). 
\item To further account for the possibility of efficient gas stripping prior to or upon infall to the group, we consider the evolution of a gas-poor galaxy (gas mass$=$$0.2$\mstar{}). 
\item The dependence on orbital orientation is addressed in three scenarios, including: Orbit:Face-on ($\theta$=90$^{\circ}$, $\phi$=0$^{\circ}$), Orbit:Edge-on ($\theta$=0$^{\circ}$, $\phi$=0$^{\circ}$) and Orbit:Retrograde ($\theta$=-45$^{\circ}$, $\phi$=-30$^{\circ}$).
\item The influence of satellite-satellite interactions (in a satellite population whose mean velocity dispersion is $\sim$150 kms$^{\rm -1}$) is established for the fiducial and B/T=0.2 models with two suites of simulations, each comprising 8 runs where the live galaxy is placed randomly on the locus defined by the group virial radius. This method accounts for the quasi-stochastic influence of the point-mass group satellites on the live galaxy model.

\end{itemize}

\begin{table} 
\centering
\caption{Summary of 21 non-fiducial models and their deviations from the fiducial; the two entries incorporating harassment are each comprised of eight simulations.}
\begin{tabular}{@{}ll@{}}
\hline
Simulation & Deviation from fiducial\\
\hline
B/T=0.2 & Initial B/T of 0.2 \\
Orbit:Edge-on & Edge-on disc w.r.t. orbit plane \\
Orbit:Face-on & Face-on disc w.r.t. orbit plane \\
Orbit:Retrograde & Retrograde w.r.t. orbit plane \\
Gas fraction:Low & Initial gas fraction of 0.2 \\
Fiducial+Harassment & Group satellites added to fiducial \\
B/T=0.2+Harassment & Group satellites added to B/T=0.2 \\
\hline
\end{tabular}
\end{table} 

\subsection{Analysis parameters}

We deduce both intrinsic and apparent specific angular momenta and ellipticities of our galaxy models. Intrinsic parameters are calculated from the 3D spatial/velocity data of all stellar particles within the half-mass radius, where the intrinsic ellipticity is established from decomposition of the inertial tensor of all stars (from which the ellipticity is 1-$c$/$a$ where $c$ and $a$ are the minor and major axes respectively).

The apparent specific angular momentum is computed according to the relation defined by \citet{emse07}:
\begin{equation}
\lambda_{\rm R} = \frac{ \sum_{i=1}^{N_{\rm b}} F_{\rm i} R_{\rm i} |V_{\rm i}| } { \sum_{i=1}^{N_{\rm b}} F_{\rm i} R_{\rm i} \sqrt{ V_{\rm i}^{\rm 2} + \sigma_{\rm i}^{\rm 2} } }
\end{equation}
where F$_{\rm i}$, R$_{\rm i}$, V$_{\rm i}$ and $\sigma_{\rm i}$ are the flux (computed here from the integrated surface brightness), radius, projected stellar velocity and stellar velocity dispersion respectively, in the $i$th spatial bin.

As in YB15, we adopt synthesis models for stellar evolution and an assumed gas-to-dust ratio of 5$\times$\ex{21}{} cm$^{\rm -2}$A$_{\rm V}$ to establish stellar luminosities from the stellar age/mass of simulation particles. Surface brightness distributions of the galaxy models can be therefore constructed on 2D spatial grids. Ellipticities are calculated according to the {\sc Kinemetry} method \citep{kraj06}, in which concentric ellipses are fitted to binned line-of-sight kinematic maps. 






\citet{emse07} showed that the apparent properties taken at the effective radius, r$_{\rm e}$ (enclosing half the projected light), are sufficient to characterise inner disc kinematics. From the radial profile of $\lambda_{\rm R}$ and the ellipse ellipticities expressed as a function of their mean radii, we compute $\lambda_{\rm Re}$ and $\epsilon_{\rm e}$ respectively by linear interpolation at r$_{\rm e}$. 

In general, our method in obtaining these values is consistent with those of previous studies \citep[e.g.][]{emse07, bois11, naab14, mood14}; however, we have not adopted their Voronoi binning in construction of the kinematic maps, as our simulation resolution is sufficient to provide the requisite signal-to-noise ratio up to r$_{\rm e}$ \citep[i.e. at least $\sqrt{N}$$\ge$20 per bin;][]{capp03}.


\begin{figure}
\centering
\includegraphics[width=1.\columnwidth]{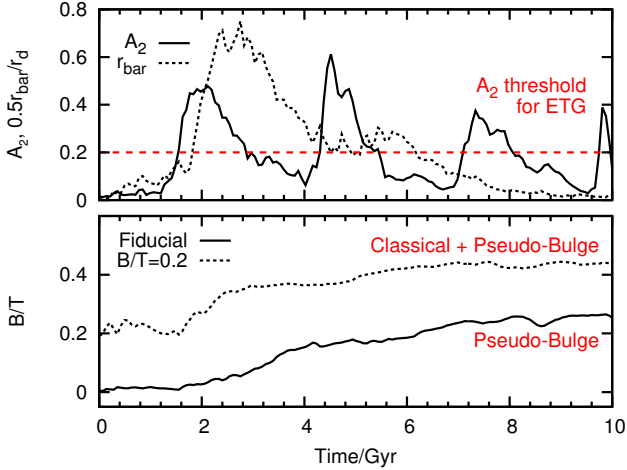}
\caption{(Top panel) Second Fourier mode amplitude (A$_{\rm 2}$, black solid line) and bar radius (r$_{\rm bar}$, normalised by stellar scalelength r$_{\rm d}$, black dashed line) as a function of time for the fiducial simulation. The red dashed line at A$_{\rm 2}=0.2$ shows our empirical threshold that separates ETGs and LTGs; (bottom panel) Bulge-to-Total ratio as a function of time for the fiducial and B/T=0.2 simulations, computed as the excess light with respect to an fitted exponential disc component. We qualify this excess as a pseudo-bulge if kinematic maps show it to be rotationally-supported.}
\end{figure}

Our mock \lre{} distributions (at a given lookback time) combine the $\lambda_{\rm Re}$ and $\epsilon_{\rm e}$ values deduced from the corresponding galaxy model data when inclined and viewed at 100 equispaced increments of $i$, where $i$ is the stellar disc inclination $0^{\circ}<i<90^{\circ}$ with respect to face-on. This approach is based on the premise that observed galaxy populations have no preferential viewing inclination. 

Furthermore, to compare this data with \atlas{} galaxies (and those of the SAMI pilot survey), whose criterion for ETGs is predicated on the absence of spiral arms, we exclude those $\lambda_{\rm Re}$ and $\epsilon_{\rm e}$ computed at timesteps where the face-on surface brightness distribution has a mean second Fourier amplitude (A$_{\rm 2}$, within 2r$_{\rm e}$) exceeding 0.2. We have confirmed that our criterion does indeed correspond to an apparent absence of spiral structure in mock B-band surface brightness imaging of our galaxy models down to $\sim$28 mag arcsec$^{\rm 2}$.

Fig. 2 compares this A$_{\rm 2}$ threshold with the computed A$_{\rm 2}$ of the fiducial model as a function of time. Also shown is the stellar bar radius, computed as in YB15. We find the model to be unstable to $m=2$ perturbations following tidal interactions at successive orbital pericentres. Subsequent gas inflow and angular momentum exchange to the bar can destroy it \citep{frie93, bour05a}, and therefore, the satellite undergoes multiple cycles between ETG and LTG classification. 

Concurrently, the bulge-to-total ratio (which we calculate from the excess light with respect to a fitted exponential disc; see YB15 for more details) grows monotonically with time (Fig. 2, bottom panel). We assert that this constitutes a pseudo-bulge for the initially bulgeless Fiducial model because the corresponding kinematic fields strongly imply its rotational-support \citep{korm04}.

\section{Results}

\begin{figure}
\centering
\includegraphics[width=1.\columnwidth]{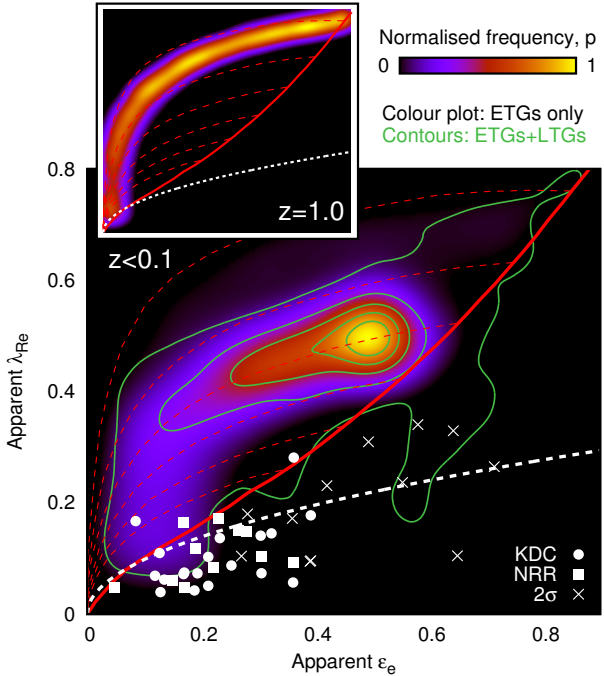}
\caption{A mock distribution of apparent specific angular momentum, $\lambda_{\rm Re}$ vs. apparent ellipticity, $\epsilon_{\rm e}$ (both measured at at r$_{\rm e}$) for the Fiducial galaxy model at \z{1.0}{} (inset panel) and $<0.1$ (main panel). The 2D probability density plot in each panel shows the normalised distribution of ETGs (as classified with A$_{\rm 2}$). In the bottom panel, this is compared with the full LTG+ETG population (green contours), together with those FRs identified in \atlas{} \citep{emse11} to appear as SRs (white symbols). The white dashed line shows an empirical boundary between SR/FRs ($\lambda_{\rm Re}=0.31\epsilon_{\rm e}^{\rm 0.5}$). The red solid line shows show an isotropic stellar system viewed edge-on would appear in this plot; the red dashed lines show how this varies with viewing inclination (where face-on lies at the origin).}
\end{figure}

Fig. 3 shows the mock probability distribution of \lre{} for our fiducial simulation at redshifts \z{1}{} and $<$0.1 (the priority coverage of the full SAMI survey). This distribution is comprised of galaxies with stellar masses in the range \ex{9-9.5}{} \msol{} due to the combined effect of secular/triggered SF and gas stripping acting on a disc with initial mass \mstar{}$=$\ex{9.25}{} \msol{}. Overlaid upon each density plot is the theoretical properties of an edge-on stellar system (solid red line) assuming a two-integral dynamical Jeans model \citep{capp07, emse11}, and its corresponding projection towards face-on (dashed lines, towards the origin). 

We find at an early epoch that our (predominantly Sc/d) satellite population is spread along along the upper-most dashed arcs (corresponding to an intrinsic $\epsilon_{\rm e}=0.7-0.8$). At lower redshifts, this distribution becomes more concentrated with a clear peak forming by $z<0.1$ at $\epsilon_{\rm e}=0.5$, $\lambda_{\rm Re}=0.5$. Fig. 2 (top panel) shows that in the intervening time the B/T ratio grows from 0 to 0.25; the corresponding (pseudo-)bulge is rotationally-supported and dominated by young stars.

An outlying proportion of the ETG population falls below an empirical criterion from \atlas{} that denotes slow rotation ($\lambda_{\rm Re}=0.31\sqrt{\epsilon_{\rm e}}$). Since we enforce a consistent definition of \lr{} based on apparent properties at r$_{\rm e}$, we expect some apparent SR classifications to arise due to a stellar bar. This bar, the size of which diminishes with time after a peak at first pericentre (Fig. 2), distorts the ellipicity metric in viewing inclinations close to face-on \citep{capp07}. 

Our autonomous ETG criterion based on A$_{\rm 2}$ filters out many more such SR classifications arising from the full LTG+ETG population (Fig. 3, green contours); this distribution shows a {\it tendril} at $\epsilon_{\rm e}=0.6, \lambda_{\rm Re}$=0.3 which lies outside the theoretical bounds for an oblate axisymmetric system, and corresponds to strong tidal features formed during the first orbital pericentre.

The capacity for group tidal interactions to induce non-equilibrium kinematics in the satellite is demonstrated here also in terms of intrinsic properties (Fig. 4). Here we find in general a close concordance between our simulation of a single fiducial galaxy model (with infall at 10 Gyr ago) and the best-fit relation for a two-integral Jeans dynamical system \citep{capp07, emse11}. This applies across a wide range of $\epsilon_{\rm e}$, albeit with some significant transient deviations at orbital pericentres (e.g. $\epsilon_{\rm e}=0.5, \lambda_{\rm Re}$=0.5). 

The existence of apparent SRs are therefore traced to transient perturbations to an otherwise persistently disc-like kinematic field \citep{kraj11, barr15}, and also projection effects including that due to bars which we note are undetected by photometry in as many as 10 percent of galaxies in the {\sc CALIFA} survey \citep{holm15}. The location of apparent SRs (at $\epsilon_{\rm e}=0.2$, $\lambda_{\rm Re}=0.1$; Fig. 3) coincides with the simulated remnants of late mergers \citep{naab14}, and observed \atlas{} galaxies whose line-of-sight velocity distributions show kinematically-decoupled components (KDCs), non-regular rotation (NRR) or double-sigma components (2$\sigma$) as opposed to pressure-supported kinematics \citep[white symbols;][]{emse11}. 

\begin{figure}
\centering
\includegraphics[width=1.\columnwidth]{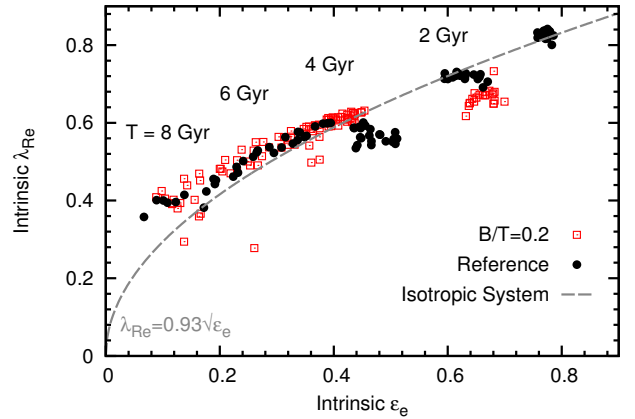}
\caption{The intrinsic evolution of the Fiducial simulation (black filled circles) and B/T=0.2 (red open squares)in the \lre{} plane; black labels give an approximate time since the start of the simulation. The approximate best-fit relation ($\lambda_{\rm Re}=0.93\epsilon_{\rm e}^{\rm 0.5}$) to the expected evolution of an isotropic stellar system is conveyed with the grey dashed line.}
\end{figure}

\subsection{Parameter dependencies}

Fig. 5 (top panel) shows how the $z<0.1$ \lre{} distribution depends on our initial galaxy model parameters. The key results are summarised as follows:
\begin{itemize}

\item Our B/T=0.2 model displays an concentration lying closer to the origin than the fiducial case. This is to be expected from an inner disc hosting a classical (pressure-supported) bulge in addition to the rotation-supported pseudo-bulge formed also in the fiducial case (with a final B/T=0.4; Fig. 2). Though not shown in this figure, we note that fewer apparent SRs are detected due in part to an inner disc less favourable for the radial ($x_{\rm 1}$) stellar orbits that constitute a stellar bar \citep{norm96}. 

\item Gas fraction:Low does not form a pseudo-bulge as efficiently as the fiducial case, where in the latter case the loss in gas angular momentum by successive tidal torques fuels bulge growth \citep{barn91}. In spite of this, the \lre{} distribution shows a small deviation towards higher $e_{\rm e}$, a result similar to that found among the simulated remnants of dissipative/collisonless mergers \citep{jess09}. Of further note is that the relatively small dissipative component allows the triggered bar (and its associated distortion of $\epsilon_{\rm e}$) to persist on order of a Hubble time causing a higher incidence of SR mis-identifications. 

\item The \lre{} distribution shows a clear dependence on the satellite orbit, a proxy to the dependence on tidal torque efficiency. Orbit:Face-on is not sufficiently perturbed to facilitate much morphological/kinematic evolution, and there exists negligible variation between prograde or retrograde orbits. In general, we find \lr{} and $\epsilon_{\rm e}$ drop sharply at orbital pericentres, as a function of time, by values that scale with ${\sim}cos(i)$ (where $i$ is galaxy inclination to its orbit). 

\end{itemize}

\subsection{Dependence on tidal harassment}

Fig. 5 (Bottom panel) shows the \lre{} distribution for Fiducial + Harassment at $z<0.1$. The addition of point-mass group satellites causes the dispersion of the highly concentrated distribution associated with our fiducial galaxy model at an intermediate orbital inclination (top panel). This dispersion, while largely remaining within theoretical limits (red lines), is due to the quasi-stochasticity of satellite-satellite interaction impact parameters and inclinations. As in the case of simulated minor mergers wherein kinematically-distinct stellar components often arise \citep{bois11}, the proportion of SR misidentifications is more significant than in our fiducial case (Fig. 3).

\begin{figure}
\centering
\includegraphics[width=1.\columnwidth]{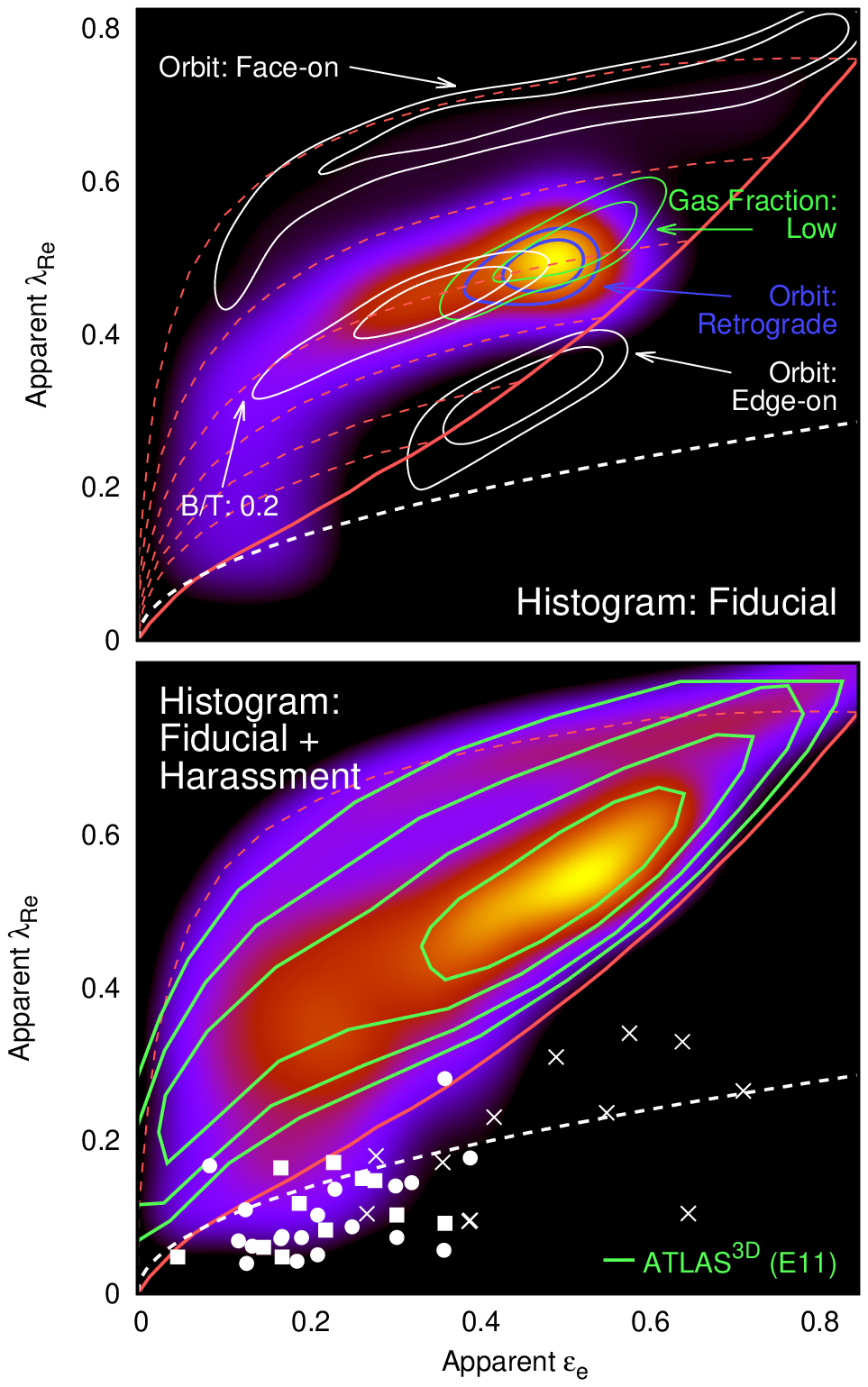}
\caption{(Top panel) As Fig. 3, but where the \lre{} distribution of the fiducial simulation at $z<0.1$ (2D probability density plot) is compared with simulations commencing with differing galaxy model conditions (contours shown at $p=0.75$ and 0.875); (bottom panel) As Fig. 3, but showing the ensemble \lre{} distribution of Harassment at $z<0.1$. Green contours represent the mock distribution of 50,000 FR ETGs \citep{emse11} based on the Monte-Carlo method of \citet{capp07}.}
\end{figure}

\section{Discussion and conclusions}






Although $\lambda_{\rm Re}$ has already proven a useful metric for distinguishing the formation paths of ETGs, galaxy evolution on the \lre{} plane remains a complex, non-linear process within a $\Lambda$CDM cosmology \citep{naab14}. In this study, we demonstrate how it is instructive to consider the global \lre{} distribution of many galaxies, analysed in a consistent manner and including projection effects. In environments like groups, a target for the upcoming SAMI survey among others, this method can overcome the non-monotonic evolution of individual satellites. Instead, we can describe the effective redshift-evolution of the group itself, which we euphemistically describe as undergoing a process of {\it global warming}. 

This demonstration was performed with a suite of idealised simulations derived from those of \citet{yozi15b} and thus shares the same limitations, including a fixed spherical group potential. We argue that since the intrinsic evolution of our satellites generally follow that of an oblate FR (Fig. 4), they will tend towards a similar \lre{} distribution albeit at a rate determined by orbital parameters. This particular evolution is predicated on an absence of major mergers, which we deem rare for the \mstar{}$=$\ex{9}{} \msol{} satellites considered here, in accordance with the ILLUSTRIS simulation which predicts $\sim$0.1 such events per satellite since \z{2}{} \citep{rodr15}. 

Significant minor mergers (e.g. mass ratios of 1:10 to 1:6) will be more frequent, however; previous numerical studies \citep[e.g.][]{bour05b} have suggested these events yield remnants analogous to S0s, and comparable therefore to the FR ETGs produced by our methodology (where bulge growth is likewise promoted by in-situ SF). However, this parallelism should be explicitly challenged in a future study. 

Hydrodynamical interactions between the satellite and the intra-group medium are also ignored; our results and those of \citet{bois11} have suggested that the extant gas fraction has only a small impact on \lre{}, although smooth accretion has been shown to raise \lr{} if suitably aligned with the disc \citep{chri15}. We do not expect, however, the environmental mechanism of ram pressure stripping will significantly influence a kinematics-driven evolution of \lre{}; in fact, this can be exploited in efforts to discern galaxy formation mechanisms \citep[e.g. S0s, see also][]{quer15}.

Our models also lack an explicit model of galactic outflow, which recent cosmological simulations demonstrate as a key actor in regulating angular momentum \citep{gene15}. Our preliminary simulations suggest the effect of stronger feedback in our model (applied with a longer adiabatic timescale, e.g. $\tau_{\rm SN}={\rm 10}^{\rm 6}$ yr) merely controls the rate of \lre{} evolution rather than its shape, the latter being the focus of this present study. 

Lastly, several studies have noted that SR formation is highly resolution dependent \citep{bois10}, insofar as convergence requires a baryonic mass resolutions of at least $\sim$\ex{4}{} \msol{}; indeed, the merger simulations of \citet{mood14} adopt this threshold and find noise fluctuations cause $\lambda_{\rm Re}$ to deviate by only $\sim$0.01. Since our simulations also lie at this threshold, we maintain our claim that the group mechanisms modelled here produce only FR ETGs.

\begin{figure}
\centering
\includegraphics[width=1.\columnwidth]{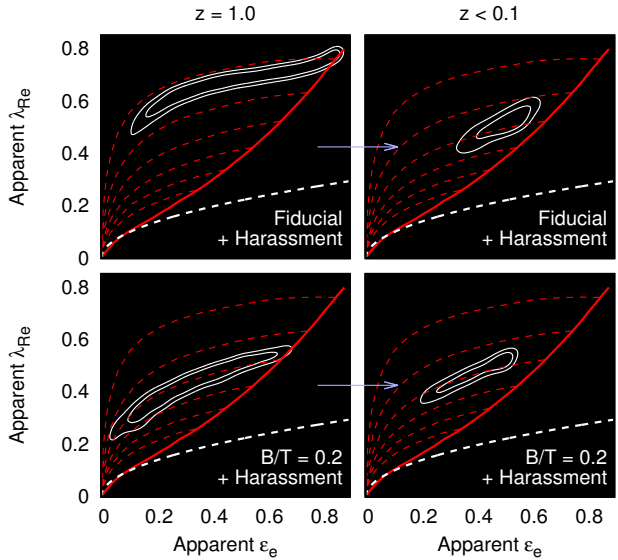}
\caption{As Fig. 3, but for mock \lre{} distributions derived from Fiducial + Harassment and B/T=0.2 + Harassment at \z{1.0}{} and $<0.1$. White contours are chosen at $p$=0.7 and 0.85 to illustrate the distinct evolutions associated with disc-dominated FR ETGs (B/T=0 and B/T=0.2) over $\sim$8 Gyr.}
\end{figure}

In an attempt to verify our idealised numerical methodology, we compare our results in Fig. 5 (bottom panel) to the mock distribution of 50,000 FR ETGs galaxies obtained by \citet{emse11}. Their Monte-Carlo method follows from a linear relation found between intrinsic $\epsilon$ and anisotropy among {\sc SAURON} galaxies \citep{capp07}, for which the associated scatter is modelled as a Gaussian \citep[see also][]{weij14}. The resulting mock distribution (green contours) shows qualitative agreement with both the \atlas{} sample of FRs (which we note is drawn from volume-limited surveys and therefore comprises satellite and central galaxies), and our $z<0.1$ snapshot of Fiducial + Harassment. 


Specifically, Fiducial + Harasment reproduces the spread of \atlas{} galaxies and their concentration around $e_{\rm e}=0.5, \lambda_{\rm Re}=0.5$. Since the majority of the \atlas{} galaxies were identified by morphological type $T$ as S0s, we can claim therefore that our simulations capture the fundamental mechanisms that yield a S0 morphology. Fig. 6 (bottom panels) illustrates how those FR ETGs galaxies hosting an isotropic bulge (classically the remnant of merger events) will appear to be more concentrated towards the origin; this \lre{} distribution is therefore qualitatively consistent with that of the pilot-SAMI ETGs identified as cluster members by \citet{foga15}. 






Of course, a comparison to the \atlas{}/SAURON and pilot-SAMI catalogs is not strictly correct (given their generally more massive stellar masses with respect to that represented by our models) and thus we look forward to the full SAMI survey for future comparisons.

However, this cursory verification of our low-redshift mock \lre{} distributions of ETG FRs allows us to propose, firstly, that the observed data (for a given stellar mass) represents the superposition of \lre{} distributions specific to particular morphological groups. \citet{capp13b} have already shown the bulge fraction can be more reliably inferred from kinematics than bulge-disc decompositions; our broader assertion is that the decomposition of \lre{} will in principle provide insight as to the evolutionary processes involved in those satellite populations \citep[e.g. pre-processing, and the relative role of classical and pseudo-bulge building mechanisms;][]{delu12, wetz13}. 

This concept complements that of the fundamental/mass plane (FP), a tight scaling relation for ETGs, which the SAMI survey has already proven capable in its construction \citep{scot15}. Unlike our concept, however, it is not conclusive if properties of the FP (e.g. slope, offset) vary with environment \citep[e.g.][]{hou15}, and indeed its utility as a distance indicator would suffer if it did \citep{mago12}.


%

A second proposition is that the tidal-driven evolution of a satellite population drives an increasing spheroidal fraction with time, which manifests qualitatively as an increasing concentration of the \lre{} distribution. This can therefore be exploited to parametrise the {\it global warming} of the environment. While admittedly a non-trivial exercise given a cosmological assembly bias, the well-known environment-driven evolution of dwarf satellites since \z{1}{} makes them the ideal targets for this concept \citep{geha12}. Indeed, we advocate the comparison of \lre{} from high redshift IFU surveys \citep[e.g. the KMOS Redshift One Spectroscopic Survey, presently conducted to compare star forming conditions within 800 star forming galaxies at $z=1$;][]{stot16} with local counterparts to verify the redshift-dependent behaviour illustrated in Fig. 6. 

\section*{Acknowledgements}

We thank the anonymous referee for their comments which improved this paper. Part of this work was supported by the Australian Postgraduate Award Scholarship. The authors also wish to thank members of the Computational Theory Group at ICRAR for their useful comments.

\bibliographystyle{mnras}
\bibliography{bib}

\begin{thebibliography}{}
\makeatletter
\relax
\def\mn@urlcharsother{\let\do\@makeother \do\$\do\&\do\#\do\^\do\_\do\%\do\~}
\def\mn@doi{\begingroup\mn@urlcharsother \@ifnextchar [ {\mn@doi@}
  {\mn@doi@[]}}
\def\mn@doi@[#1]#2{\def\@tempa{#1}\ifx\@tempa\@empty \href
  {http://dx.doi.org/#2} {doi:#2}\else \href {http://dx.doi.org/#2} {#1}\fi
  \endgroup}
\def\mn@eprint#1#2{\mn@eprint@#1:#2::\@nil}
\def\mn@eprint@arXiv#1{\href {http://arxiv.org/abs/#1} {{\tt arXiv:#1}}}
\def\mn@eprint@dblp#1{\href {http://dblp.uni-trier.de/rec/bibtex/#1.xml}
  {dblp:#1}}
\def\mn@eprint@#1:#2:#3:#4\@nil{\def\@tempa {#1}\def\@tempb {#2}\def\@tempc
  {#3}\ifx \@tempc \@empty \let \@tempc \@tempb \let \@tempb \@tempa \fi \ifx
  \@tempb \@empty \def\@tempb {arXiv}\fi \@ifundefined
  {mn@eprint@\@tempb}{\@tempb:\@tempc}{\expandafter \expandafter \csname
  mn@eprint@\@tempb\endcsname \expandafter{\@tempc}}}

\bibitem[\protect\citeauthoryear{Barnes \& Hernquist}{Barnes \&
  Hernquist}{1991}]{barn91}
Barnes J.~E.,  Hernquist L.,  1991, ApJ, 370, L65

\bibitem[\protect\citeauthoryear{Barrera-Ballesteros
  et~al.}{Barrera-Ballesteros et~al.}{2015}]{barr15}
Barrera-Ballesteros J.~K.,  et~al., 2015, A\&A, 582, A21

\bibitem[\protect\citeauthoryear{Bekki}{Bekki}{2009}]{bekk09}
Bekki K.,  2009, MNRAS, 399, 2221

\bibitem[\protect\citeauthoryear{Bekki \& Couch}{Bekki \& Couch}{2011}]{bekk11}
Bekki K.,  Couch W.,  2011, MNRAS, 415, 1783

\bibitem[\protect\citeauthoryear{Bois et~al.}{Bois et~al.}{2010}]{bois10}
Bois M.,  et~al., 2010, MNRAS, 406, 2405

\bibitem[\protect\citeauthoryear{Bois et~al.}{Bois et~al.}{2011}]{bois11}
Bois M.,  et~al., 2011, MNRAS, 416, 1654

\bibitem[\protect\citeauthoryear{Bournard, Combes  \& Semelin}{Bournard
  et~al.}{2005a}]{bour05a}
Bournard F.,  Combes F.,   Semelin B.,  2005a, MNRAS, 364, L18

\bibitem[\protect\citeauthoryear{Bournard, Jog  \& Combes}{Bournard
  et~al.}{2005b}]{bour05b}
Bournard F.,  Jog C.~J.,   Combes F.,  2005b, 437, 69

\bibitem[\protect\citeauthoryear{Bournard, Jog  \& Combes}{Bournard
  et~al.}{2007}]{bour07}
Bournard F.,  Jog C.~J.,   Combes F.,  2007, 476, 1179

\bibitem[\protect\citeauthoryear{Cappellari \& Copin}{Cappellari \&
  Copin}{2003}]{capp03}
Cappellari M.,  Copin Y.,  2003, MNRAS, 342, 345

\bibitem[\protect\citeauthoryear{Cappellari et~al.}{Cappellari
  et~al.}{2007}]{capp07}
Cappellari M.,  et~al., 2007, MNRAS, 379, 418

\bibitem[\protect\citeauthoryear{Cappellari et~al.}{Cappellari
  et~al.}{2013a}]{capp13b}
Cappellari M.,  et~al., 2013a, MNRAS, 432, 1862

\bibitem[\protect\citeauthoryear{Cappellari et~al.}{Cappellari
  et~al.}{2013b}]{capp13}
Cappellari M.,  et~al., 2013b, ApJ, 778, L2

\bibitem[\protect\citeauthoryear{Christensen, Dav{\'e}, Governato, Pontzen,
  Brooks, Munshi, Quinn  \& Wadsley}{Christensen et~al.}{2015}]{chri15}
Christensen C.~R.,  Dav{\'e} R.,  Governato F.,  Pontzen A.,  Brooks A.,
  Munshi F.,  Quinn T.,   Wadsley J.,  2015, Submitted to ApJ; preprint
  (arXiv:1508.00007v1)

\bibitem[\protect\citeauthoryear{Croom et~al.}{Croom et~al.}{2012}]{croo12}
Croom S.~M.,  et~al., 2012, MNRAS, 421, 872

\bibitem[\protect\citeauthoryear{{De Lucia}, Weinmann, Poggianti,
  Arag{\'o}n-Salammanca  \& Zaritksy}{{De Lucia} et~al.}{2012}]{delu12}
{De Lucia} G.,  Weinmann S.,  Poggianti B.~M.,  Arag{\'o}n-Salammanca A.,
  Zaritksy D.,  2012, MNRAS, 423, 1277

\bibitem[\protect\citeauthoryear{{\VAN{Deugenio}{d'Eugenio}{d'Eugenio}},
  Houghton, Davies  \& Bont{/` a}}{{\VAN{Deugenio}{d'Eugenio}{d'Eugenio}}
  et~al.}{2015}]{deug15}
{\VAN{Deugenio}{d'Eugenio}{d'Eugenio}} F.,  Houghton R. C.~W.,  Davies R.~L.,
  Bont{/` a} E.~D.,  2015

\bibitem[\protect\citeauthoryear{Dressler}{Dressler}{1980}]{dres80}
Dressler A.,  1980, ApJ, 236, 351

\bibitem[\protect\citeauthoryear{Emsellem et~al.}{Emsellem
  et~al.}{2007}]{emse07}
Emsellem E.,  et~al., 2007, MNRAS, 379, 401

\bibitem[\protect\citeauthoryear{Emsellem et~al.}{Emsellem
  et~al.}{2011}]{emse11}
Emsellem E.,  et~al., 2011, MNRAS, 414, 888

\bibitem[\protect\citeauthoryear{Fogarty et~al.}{Fogarty et~al.}{2014}]{foga15}
Fogarty L. M.~R.,  et~al., 2014, preprint (arXiv:1509.02641)

\bibitem[\protect\citeauthoryear{Friedli \& Benz}{Friedli \&
  Benz}{1993}]{frie93}
Friedli D.,  Benz W.,  1993, A\&A, 268, 65

\bibitem[\protect\citeauthoryear{Geha, Blanton, Yan  \& Tinker}{Geha
  et~al.}{2012}]{geha12}
Geha M.,  Blanton M.~R.,  Yan R.,   Tinker J.~L.,  2012, ApJ, 757, 85

\bibitem[\protect\citeauthoryear{Genel, Fall, Hernquist, Vogelsberger, Snyder,
  Rodriguez-Gomez, Sijacki  \& Springel}{Genel et~al.}{2015}]{gene15}
Genel S.,  Fall S.~M.,  Hernquist L.,  Vogelsberger M.,  Snyder G.~F.,
  Rodriguez-Gomez V.,  Sijacki D.,   Springel V.,  2015, ApJ, 804, L40

\bibitem[\protect\citeauthoryear{Holmes et~al.}{Holmes et~al.}{2015}]{holm15}
Holmes L.,  et~al., 2015, MNRAS, 451, 4397

\bibitem[\protect\citeauthoryear{Hou \& Wang}{Hou \& Wang}{2015}]{hou15}
Hou L.,  Wang Y.,  2015, Research in Astronomy and Astrophysics, 15, 651

\bibitem[\protect\citeauthoryear{Huang, Haynes, Giovanelli, Brinchmann,
  Stierwalt  \& Neff}{Huang et~al.}{2012}]{huan12}
Huang S.,  Haynes M.~P.,  Giovanelli R.,  Brinchmann J.,  Stierwalt S.,   Neff
  S.~G.,  2012, AJ, 143, 133

\bibitem[\protect\citeauthoryear{Ichikawa, Kajisawa  \& Akhlaghi}{Ichikawa
  et~al.}{2012}]{ichi12}
Ichikawa T.,  Kajisawa M.,   Akhlaghi M.,  2012, MNRAS, 422, 1014

\bibitem[\protect\citeauthoryear{Jesseit, Cappellari, Naab, Emsellem  \&
  Burkert}{Jesseit et~al.}{2009}]{jess09}
Jesseit R.,  Cappellari M.,  Naab T.,  Emsellem E.,   Burkert A.,  2009, MNRAS,
  397, 1202

\bibitem[\protect\citeauthoryear{Just, Zaritsky, Sand, Desai  \& Rudnick}{Just
  et~al.}{2010}]{just10}
Just D.~W.,  Zaritsky D.,  Sand D.~J.,  Desai V.,   Rudnick G.,  2010, ApJ,
  711, 192

\bibitem[\protect\citeauthoryear{Khochfar \& Silk}{Khochfar \&
  Silk}{2006}]{khoc06}
Khochfar S.,  Silk J.,  2006, MNRAS, 370, 902

\bibitem[\protect\citeauthoryear{Kormendy \& Kennicutt}{Kormendy \&
  Kennicutt}{2004}]{korm04}
Kormendy J.,  Kennicutt R.~C.,  2004, ARA\&A, 42, 603

\bibitem[\protect\citeauthoryear{Krajnovi{\' c}, Cappellari, {Tim de Zeeuw}  \&
  Copin}{Krajnovi{\' c} et~al.}{2006}]{kraj06}
Krajnovi{\' c} D.,  Cappellari M.,  {Tim de Zeeuw} P.,   Copin Y.,  2006,
  MNRAS, 366, 787

\bibitem[\protect\citeauthoryear{Krajnovi{\' c} et~al.}{Krajnovi{\' c}
  et~al.}{2011}]{kraj11}
Krajnovi{\' c} D.,  et~al., 2011, MNRAS, 414, 2923

\bibitem[\protect\citeauthoryear{Kravtsov}{Kravtsov}{2013}]{krav13}
Kravtsov A.~V.,  2013, ApJL, 764, 31

\bibitem[\protect\citeauthoryear{Magoulas et~al.}{Magoulas
  et~al.}{2012}]{mago12}
Magoulas C.,  et~al., 2012, MNRAS, 427, 245

\bibitem[\protect\citeauthoryear{Miller, Ellis, Newman  \& Benson}{Miller
  et~al.}{2014}]{mill14}
Miller S.~H.,  Ellis R.~S.,  Newman A.~B.,   Benson A.,  2014, ApJ, 782, 2

\bibitem[\protect\citeauthoryear{Moody, Romanowsky, Cox, Novak  \&
  Primack}{Moody et~al.}{2014}]{mood14}
Moody C.~E.,  Romanowsky A.~J.,  Cox T.~J.,  Novak G.~S.,   Primack J.~R.,
  2014, MNRAS, 444, 1475

\bibitem[\protect\citeauthoryear{Munoz-Cuartas, Maccio, Gottlober  \&
  Dutton}{Munoz-Cuartas et~al.}{2011}]{muno11}
Munoz-Cuartas J.~C.,  Maccio A.~V.,  Gottlober S.,   Dutton A.~A.,  2011,
  MNRAS, 584, 94

\bibitem[\protect\citeauthoryear{Naab et~al.}{Naab et~al.}{2014}]{naab14}
Naab T.,  et~al., 2014, MNRAS, 444, 3357

\bibitem[\protect\citeauthoryear{Norman, Sellwood  \& Hasan}{Norman
  et~al.}{1996}]{norm96}
Norman C.~A.,  Sellwood J.~A.,   Hasan H.,  1996, ApJ, 462, 114

\bibitem[\protect\citeauthoryear{Obreshkow \& Glazebrook}{Obreshkow \&
  Glazebrook}{2014}]{obre14}
Obreshkow D.,  Glazebrook K.,  2014, ApJ, 784, 26

\bibitem[\protect\citeauthoryear{Popesso et~al.}{Popesso et~al.}{2015}]{pope15}
Popesso P.,  et~al., 2015, 579, 132

\bibitem[\protect\citeauthoryear{Querejeta et~al.}{Querejeta
  et~al.}{2015}]{quer15}
Querejeta M.,  et~al., 2015, preprint (arXiv:1506.00640)

\bibitem[\protect\citeauthoryear{Rodriguez-Gomez et~al.}{Rodriguez-Gomez
  et~al.}{2015}]{rodr15}
Rodriguez-Gomez V.,  et~al., 2015, MNRAS, 449, 49

\bibitem[\protect\citeauthoryear{Scott et~al.}{Scott et~al.}{2015}]{scot15}
Scott N.,  et~al., 2015, preprint (arXiv:1505.04354)

\bibitem[\protect\citeauthoryear{Stott et~al.}{Stott et~al.}{2016}]{stot16}
Stott J.~P.,  et~al., 2016, preprint (arXiv:1601.03400)

\bibitem[\protect\citeauthoryear{Villalobos, {De Lucia}, Borgani  \&
  Murante}{Villalobos et~al.}{2012}]{vill12}
Villalobos A.,  {De Lucia} G.,  Borgani S.,   Murante G.,  2012, MNRAS, 424,
  2401

\bibitem[\protect\citeauthoryear{Weijmans et~al.}{Weijmans
  et~al.}{2014}]{weij14}
Weijmans A.-M.,  et~al., 2014, MNRAS, 444, 3340

\bibitem[\protect\citeauthoryear{Weinzirl, Jogee, Khochfar, Burkert  \&
  Kormendy}{Weinzirl et~al.}{2009}]{wein09}
Weinzirl T.,  Jogee S.,  Khochfar S.,  Burkert A.,   Kormendy J.,  2009, ApJ,
  696, 411

\bibitem[\protect\citeauthoryear{Wetzel, Tinker, Conroy  \& {van den
  Bosch}}{Wetzel et~al.}{2013}]{wetz13}
Wetzel A.~R.,  Tinker J.~L.,  Conroy C.,   {van den Bosch} F.~C.,  2013, MNRAS,
  432, 336

\bibitem[\protect\citeauthoryear{Wilman, Oemler, Mulchaey, McGee, Balogh  \&
  Bower}{Wilman et~al.}{2009}]{wilm09}
Wilman D.~J.,  Oemler A.,  Mulchaey J.~S.,  McGee S.~L.,  Balogh M.~L.,   Bower
  R.~G.,  2009, ApJ, 692, 298

\bibitem[\protect\citeauthoryear{Yozin \& Bekki}{Yozin \&
  Bekki}{2015}]{yozi15b}
Yozin C.,  Bekki K.,  2015, MNRAS, 453, 14

\makeatother
\end{thebibliography}

\bsp
\label{lastpage}
\end{document}